\documentclass[twocolumn]{aastex631}
\usepackage{mathtools} 
\usepackage{csvsimple}
\usepackage{longtable}
\usepackage{footmisc}
\renewcommand{\thefootnote}{\arabic{footnote}}

\shorttitle{Householder \& Weiss et al.}
\shortauthors{Householder \& Weiss et al.}
\graphicspath{{./}{figures/}}
\usepackage{orcidlink}

\begin{document}

\title{Investigating the Atmospheric Mass Loss of the Kepler-105 Planets Straddling the Radius Gap}
\author[0000-0002-5812-3236]{Aaron Householder}
\altaffiliation{MIT Dean of Science Fellow}
\affiliation{Department of Earth, Atmospheric and Planetary Sciences, Massachusetts Institute of Technology, Cambridge, MA 02139, USA}
\affiliation{Department of Astronomy, Yale University,
52 Hillhouse, New Haven, CT 06511, USA}

\author[0000-0002-3725-3058]{Lauren M. Weiss}
\affiliation{Department of Physics and Astronomy, University of Notre Dame, Notre Dame, IN 46556, USA}

\author[0000-0002-4856-7837]{James E. Owen}
\affil{Astrophysics Group, Department of Physics, Imperial College London, Prince Consort Rd, London SW7 2AZ, UK}

\author[0000-0002-0531-1073]{Howard Isaacson}
\affiliation{Department of Astronomy,  University of California Berkeley, Berkeley CA 94720, USA}
\affiliation{Centre for Astrophysics, University of Southern Queensland, Toowoomba, QLD, Australia}

\author[0000-0001-8638-0320]{Andrew W. Howard}
\affiliation{Department of Astronomy, California Institute of Technology, Pasadena, CA 91125, USA}

\author[0000-0003-3750-0183]{Daniel Fabrycky}
\affiliation{Dept. of Astronomy \& Astrophysics, University of Chicago, 5640 S. Ellis Ave., Chicago, IL 60637}

\author[0000-0003-0638-3455]{Leslie A. Rogers}
\affiliation{Department of Astronomy and Astrophysics, University of Chicago, Chicago, IL 60637, USA}

\author[0000-0002-0298-8089]{Hilke E. Schlichting}
\affiliation{Department of Earth, Planetary, and Space Sciences, The University of California, Los Angeles, 595 Charles E. Young Drive East, Los Angeles, CA 90095, USA}
\author[0000-0003-3504-5316]{Benjamin J.\ Fulton}
\affiliation{Cahill Center for Astronomy $\&$ Astrophysics, California Institute of Technology, Pasadena, CA 91125, USA}
\affiliation{IPAC-NASA Exoplanet Science Institute, Pasadena, CA 91125, USA}

\author[0000-0003-0967-2893]{Erik A. Petigura}
\affil{Department of Physics \& Astronomy, University of California Los Angeles, Los Angeles, CA 90095, USA}

\author[0000-0002-8965-3969]{Steven Giacalone}
\affiliation{Department of Astronomy,  University of California Berkeley, Berkeley CA 94720, USA}
\affiliation{Department of Astronomy, California Institute of Technology, Pasadena, CA 91125, USA}

\author[0000-0001-8898-8284]{Joseph M. Akana Murphy}
\altaffiliation{NSF Graduate Research Fellow}
\affiliation{Department of Astronomy and Astrophysics, University of California, Santa Cruz, CA 95064, USA}

\author[0000-0001-7708-2364]{Corey Beard}
\altaffiliation{NASA FINESST Fellow}
\affiliation{Department of Physics \& Astronomy, The University of California, Irvine, Irvine, CA 92697, USA}

\author[0000-0003-1125-2564]{Ashley Chontos}
\affiliation{Department of Astrophysical Sciences, Princeton University, 4 Ivy Lane, Princeton, NJ 08540, USA}
\affiliation{Institute for Astronomy, University of Hawai’i, 2680 Woodlawn Drive, Honolulu, HI 96822 USA}

\author[0000-0002-8958-0683]{Fei Dai}
\altaffiliation{NASA Sagan Fellow}
\affiliation{Division of Geological and Planetary Sciences,
1200 E California Blvd, Pasadena, CA, 91125, USA}
\affiliation{Department of Astronomy, California Institute of Technology, Pasadena, CA 91125, USA}

\author[0000-0002-4290-6826]{Judah Van Zandt}
\affiliation{Department of Physics \& Astronomy, University of California Los Angeles, Los Angeles, CA 90095, USA}

\author[0000-0001-8342-7736]{Jack Lubin}
\affiliation{Department of Physics \& Astronomy, University of California Irvine, Irvine, CA 92697, USA}

\author[0000-0002-7670-670X]{Malena Rice}
\affiliation{Department of Astronomy, Yale University,
52 Hillhouse, New Haven, CT 06511, USA}

\author[0000-0001-7047-8681]{Alex S. Polanski} 
\affiliation{Department of Physics and Astronomy, University of Kansas, Lawrence, KS 66045, USA}

\author[0000-0002-4297-5506]{Paul Dalba}
\affiliation{Department of Earth \& Planetary Sciences, University of California Riverside, 900 University Ave, Riverside, CA 92521, USA}

\author[0000-0002-3199-2888]{Sarah Blunt}
\affiliation{Department of Astronomy, California Institute of Technology, Pasadena, CA 91125, USA}

\author[0000-0002-1845-2617]{Emma V. Turtelboom}\affiliation{Department of Astronomy, 501 Campbell Hall, University of California, Berkeley, CA 94720, USA}

\author[0000-0003-3856-3143]{Ryan Rubenzahl}
\altaffiliation{NSF Graduate Research Fellow}
\affiliation{Department of Astronomy, California Institute of Technology, Pasadena, CA 91125, USA} 

\author[0000-0002-4480-310X]{Casey Brinkman}
\affiliation{Institute for Astronomy, University of Hawai’i, 2680 Woodlawn Drive, Honolulu, HI 96822 USA} 

\begin{abstract}
An intriguing pattern among exoplanets is the lack of detected planets between approximately $1.5$ R$_\oplus$ and $2.0$ R$_\oplus$. One proposed explanation for this ``radius gap'' is the photoevaporation of planetary atmospheres, a theory that can be tested by studying individual planetary systems. Kepler-105 is an ideal system for such testing due to the ordering and sizes of its planets. Kepler-105 is a sun-like star that hosts two planets straddling the radius gap in a rare architecture with the larger planet closer to the host star ($R_b = 2.53\pm0.07$ R$_\oplus$, $P_b = 5.41$ days, $R_c = 1.44\pm0.04$ R$_\oplus$, $P_c = 7.13$ days).  If photoevaporation sculpted the atmospheres of these planets, then Kepler-105b would need to be much more massive than Kepler-105c to retain its atmosphere, given its closer proximity to the host star. To test this hypothesis, we simultaneously analyzed radial velocities (RVs) and transit timing variations (TTVs) of the Kepler-105 system, measuring disparate masses of $M_b = 10.8\pm2.3$ M$_\oplus$ ($ \rho_b = 3.68 \pm 0.84$ g cm$^{-3}$) and $M_c = 5.6\pm1.2$ M$_\oplus $ ($\rho_c = 10.4 \pm 2.39 $ g cm$^{-3}$). Based on these masses, the difference in gas envelope content of the Kepler-105 planets could be entirely due to photoevaporation (in 76\% of scenarios), although other mechanisms like core-powered mass loss could have played a role for some planet albedos.
\bigskip
\smallskip
\end{abstract}

\section{Introduction} 
In one of the most significant exoplanet discoveries in recent years, \citet{2017AJ....154..109F} identified a gap in the occurrence rate of exoplanets between approximately $1.5$ R$_\oplus$ and $2.0$ R$_\oplus$. Various theories have been proposed to explain this ``radius gap,'' two of which are particularly prominent: core-powered mass loss \citep{2018MNRAS.476..759G, 2019MNRAS.487...24G} and photoevaporation \citep{2017ApJ...847...29O}. As planets form and accrete gas and dust from the protoplanetary disk, they can become surrounded by a gaseous envelope. However, this primordial envelope can be removed. Core-powered mass loss facilitates the loss of planetary atmospheres due to the cooling luminosity of a planet's core. X-ray and ultraviolet (XUV) radiation from the host star can also drive atmospheric mass loss via photoevaporation, where the XUV radiation ionizes and heats up the gas in the planetary atmosphere, causing it to escape into space. Both of these processes can cause planets to lose a substantial amount of their gas envelopes, leading to a significant reduction in their overall radii. Theoretical models suggest that planets within the radius gap ($1.5$ R$_\oplus - 2.0 $ R$_\oplus$) lose their gas envelopes on short timescales, leading to a reduction in their radii, whereas planets larger than the gap have much longer timescales for atmospheric mass-loss \citep{2013ApJ...775..105O,2013ApJ...776....2L,2017ApJ...847...29O, 2020A&A...638A..52M}.  Furthermore, planets smaller than $1.5$ R$_\oplus$ that are close to their stars are typically thought to be rocky, in which case they have no atmospheres left to lose \citep{2014ApJ...783L...6W, 2015ApJ...801...41R}.

The explanation of the radius gap is sometimes framed as a binary choice between core-powered mass loss and photoevaporation. However, such a simplistic interpretation likely does not encompass the full complexity of atmospheric mass-loss for the planets in this regime. A more nuanced approach to understanding the radius gap likely involves a combination of both core-powered mass loss and photoevaporation, each playing a role in sculpting planetary architectures. Therefore, instead of seeking a definitive answer to which theory explains the radius gap, it is more prudent to explore how each mechanism contributes to shaping different types of planetary architectures. Studying individual planetary systems offers a unique advantage in this context. By focusing on planets that share the same host star properties (i.e. mass, radius, temperature, age, XUV radiation history, and metallicity), we can eliminate a multitude of confounding factors that limit broader population studies. Thus, individual planetary systems provide us with a more robust testbed for examining theories such as photoevaporation and core-powered mass loss.

Kepler-105 is a system that serves as an excellent natural laboratory for investigating the role of photoevaporation in sculpting planetary architectures. In the case of photoevaporation-driven atmospheric mass loss, the time-integrated XUV radiation that a planet receives affects how much atmosphere the planet loses. Thus, the best systems for testing photoevaporation have an unusual architecture in which a gas-rich sub-Neptune is interior to a rocky planet \citep{2020MNRAS.491.5287O}.  Kepler-105 has two confirmed planets that follow this architecture: a sub-Neptune ($R_b = 2.53 \pm 0.07 R_\oplus$) with a period of $5.41$ days and a super-Earth ($R_c = 1.44 \pm 0.04 R_\oplus$) with a period of 7.13 days \citep{2018AJ....156..264F}. Based on previous mass measurements of planets with similar sizes, we expect Kepler-105c to have a predominantly rocky composition, while Kepler-105b likely has a significant gaseous envelope. If their compositions are typical for their sizes,  it is unclear how the inner planet, which likely received more XUV flux from the host star, managed to retain a significant gaseous envelope while the smaller outer planet did not. 

A similar problem was posed for Kepler-36, a benchmark system that played an important role in developing radius valley predictions and photoevaporation models \citep{2012Sci...337..556C,2013ApJ...776....2L, 2020MNRAS.491.5287O}. Kepler-36 hosts two confirmed planets near the 6:7 mean motion resonance (MMR), with a Neptune-sized planet exterior to a super-Earth. The Neptune-sized planet was found to possess a much more massive core, thereby making it more likely to retain a gaseous envelope \citep{2013ApJ...776....2L}. These findings prompted us to explore a similar scenario for the Kepler-105 system. If Kepler-105b and Kepler-105c formed in situ, then Kepler-105b would receive $\sim 44\%$ more cumulative XUV radiation than Kepler-105c. Thus, similar to the Kepler-36 system, the mass of Kepler-105b must be substantially larger than that of Kepler-105c for the latter to have lost its envelope due to photoevaporation while the former retained it. 

To test this hypothesis, we first analyzed TTVs (Section \ref{Section2}) and RVs (Section \ref{Section3}) to measure the masses of both planets. We then jointly modeled the TTVs and RVs to refine these mass measurements in Section \ref{Section4} and explored potential planetary compositions (Section \ref{Section5}). Section \ref{Section6} compares our measured masses with numerical predictions from \texttt{EvapMass} \citep{2020ascl.soft11015O}. This allows us to assess the viability of photoevaporation to explain the observed difference in gas content between Kepler-105b and Kepler-105c. We also explore core-powered mass loss as an alternative mechanism to explain the different gas compositions of these planets. Finally, in Section \ref{Section7}, we provide a summary of our findings and outline potential avenues for future research related to Kepler-105. 

\section{Mass From TTVs}
\label{Section2}

TTVs are variations in the orbital period of a planet caused by the gravitational influence of other objects in the same system such as planets or moons. Since the amplitude of the TTV of a planet depends on the mass of the companion \citep{2012ApJ...761..122L}, we can use TTVs to measure the masses of Kepler-105b and Kepler-105c, which are interior to the 4:3 MMR \citep{2018AJ....156..264F}. We analyzed 246 transit times for Kepler-105b and 179 transit times for Kepler-105c from Q1-Q17 short and long cadence data from the \textit{Kepler Space Telescope} (\textit{priv. communication Jason Rowe, 2022}, based on \citealt{Rowe2015}), shown in Figure \ref{Fig1} with a linear ephemeris subtracted. To model the TTVs, we used \texttt{TTVFaster} \citep{2016ascl.soft04012A}, which uses perturbation theory to model all terms to first-order in eccentricity. This semi-analytic approach has been demonstrated to produce accurate results for planets that are low-mass, low-eccentricity and not too deep within resonance \citep{2016ascl.soft04012A}, such as Kepler-105b and Kepler-105c. To find the best fit to the transit times of Kepler-105b and Kepler-105c using \texttt{TTVFaster}, we maximized the following log-likelihood function:
 
\begin{equation}
\log(\mathcal{L}) = -0.5 \sum_{i} \frac{\left(\mathrm{TT}_{i}-\mathrm{TT}_{m , i}\right)^{2}}{\sigma_{\mathrm{TT}, i}^{2}}
\end{equation} 
where $\mathrm{TT}_{i}$ and $\mathrm{TT}_{m , i}$ are the observed and model-predicted transit times for the \(i\)-th observation, respectively, and $\sigma_{\mathrm{TT}, i}$ is the observational uncertainty for that transit. $\sum_{i}$ indicates that we sum over all observed transits for both Kepler-105b and Kepler-105c.

To explore various solutions in our parameter space we used $\texttt{emcee}$ \citep{2013PASP..125..306F}: a Python package that runs a Markov Chain Monte Carlo (MCMC) algorithm with an affine-invariant ensemble sampler \citep{Goodman2010}. We varied the masses ($M$), orbital periods ($P$), $\sqrt{e }$ $\mathrm{cos} \omega_p$, $ \sqrt{e }$ $   \mathrm{sin} \omega_p $, and the initial times of transit ($t_0$). We re-parameterized $e$ and $\omega$ in this way to mitigate against an artificial build up of eccentricities near zero due to the boundary condition at $e=0$ \citep{2013PASP..125...83E}. We allowed stellar mass to vary as well, using a Gaussian prior of $0.99 \pm{0.03 M_\odot}$ based on previous stellar characterization \citep{2018AJ....156..264F}. Since the planets in Kepler-105 have very close orbits, we used a Hill stability prior to prevent orbit crossing: 

\begin{figure}%
    \centering
    \includegraphics[width=0.5\textwidth]{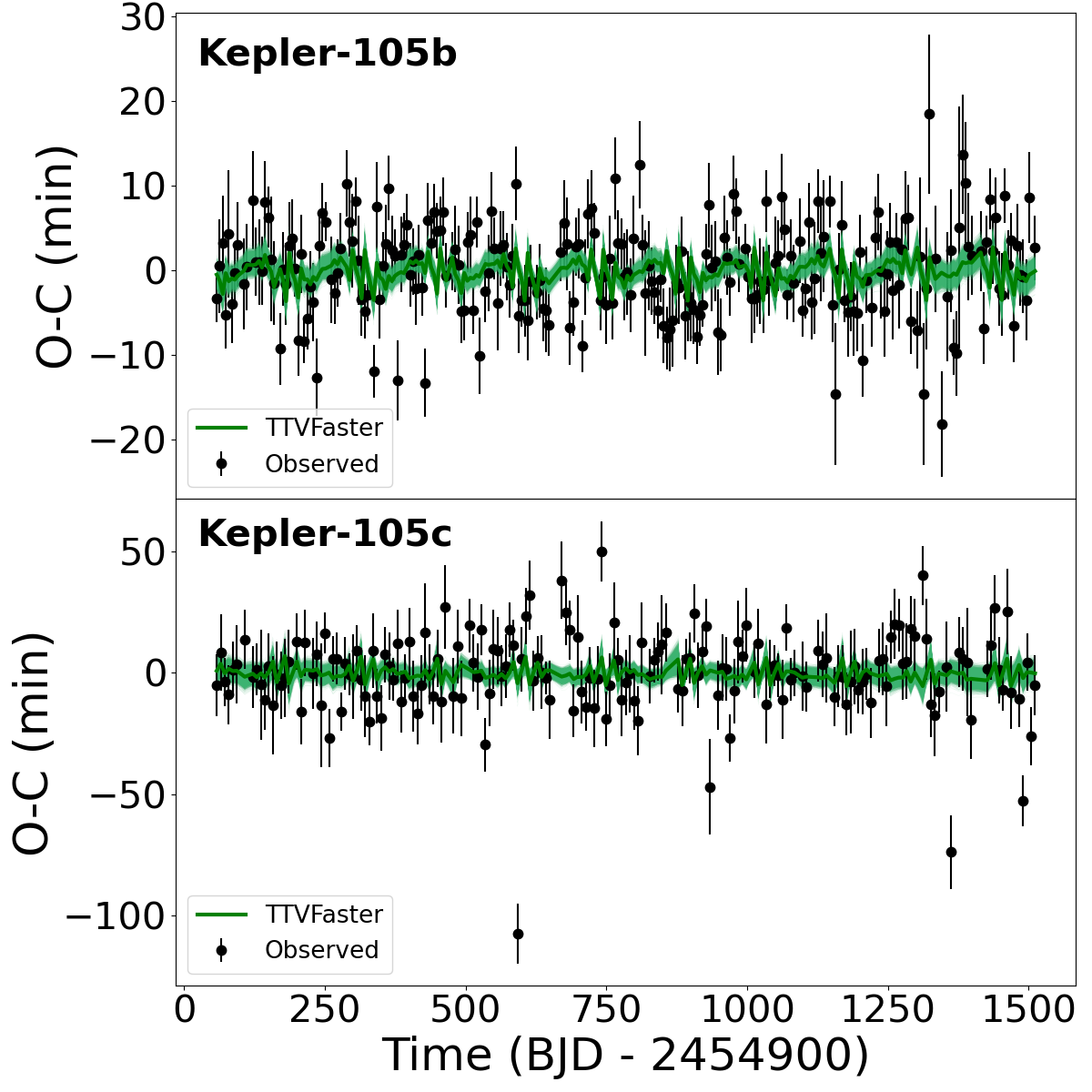}
    \caption{Observed transit times minus a linear ephemeris (black) for Kepler-105b (top) and Kepler-105c (bottom). The plot also includes the best fit \texttt{TTVFaster} \citep{2016ascl.soft04012A} solution to the TTVs (dark green) as well as the 1$\sigma$ confidence intervals from our model (light green). Our TTV model is strongly preferred to a linear ephemeris ($\Delta{AIC} = -27$, where $AIC$ is the Akaike Information Criterion,  \citealt{Akaike1974}), indicating the presence of dynamical perturbations affecting the transit times. Based on the TTVs alone, we detected Kepler-105c with $4$$\sigma$ confidence ($5.9 \pm 1.4$ M$_\oplus$) and Kepler-105b  with 2$\sigma$ confidence ($9.3^{+4.9}_{-4.6}$ M$_\oplus $). Furthermore, we ran two additional MCMC runs where the mass of Kepler-105c was constrained to be either $ \leq 3.1$ M$_\oplus$ or $\geq 8.7$ M$_\oplus$. Our best-fit \texttt{TTVFaster} analytic solution was strongly preferred over these MCMC models ($\Delta AIC = -11 $ for both cases). }
    \label{Fig1}
    \end{figure}

\begin{equation}
\label{eq2}
\begin{aligned}
& a_c \left(1-e_c\right)>a_b \ \left(1+e_b\right) + \\
& \max \left(a_b \ \left(1-e_b\right) \ \left(\frac{M_b}{3 M_{*}}\right)^{\frac{1}{3}},\right. \\
&\left.a_c \ \left(1-e_c\right) \ \left(\frac{M_c}{3 M_{*}}\right)^{\frac{1}{3}}\right)
\end{aligned}
\end{equation}
where $a$ represents the semi-major axes for Kepler-105b and Kepler-105c (denoted with subscripts b and c) and $M_*$ represents the stellar mass. Gaussian priors were placed on $P$ and $t_0$ and uniform priors were placed on $e$, $\omega$ and $M$ (see Table \ref{table1}). We determined $a$ in Equation \ref{eq2} using Kepler's Third Law from $P$ and $M_*$. We also fixed the orbital inclinations of the planets in an edge-on configuration ($i=90^\circ$).  This is because $\texttt{TTVFaster}$ assumes coplanar orbits for each planet since the amplitude of TTVs scales with mutual inclination to second-order \citep{2012ApJ...761..122L}. With this set-up, we ran the MCMC until convergence, discarding the first $10^5$ steps as burn-in. To check for convergence, we used the potential scale reduction factor (PSRF, \citealt{1992StaSc...7..457G}), requiring each parameter in our model to have a PSRF less than 1.01. 

Our MCMC analysis of the TTVs yielded a mass of $9.3^{+4.9}_{-4.6}$ M$_\oplus $ for Kepler-105b and a mass of  $5.9 \pm 1.4$ M$_\oplus $ for Kepler-105c. These findings place strong constraints on the mass of Kepler-105c (4$\sigma$), but not Kepler-105b (only 2$\sigma$). This outcome is consistent with previous TTV analyses of the system (\citealt{2017AJ....154....5H}; \citealt{2016ApJ...820...39J}). However, this may seem surprising, given that Kepler-105b is more massive and should theoretically induce larger and more easily detectable TTVs than Kepler-105c. The explanation for this is two-fold. Firstly, the mass error for Kepler-105c scales with the transit midpoint error of the larger Kepler-105b. Secondly, Kepler-105b produces a deeper transit, making it easier to precisely measure the midpoint of each transit. Thus, the higher precision in transit midpoint measurements of Kepler-105b leads to better constraints on the mass of Kepler-105c. By the same logic, the mass of Kepler-105b is not as well determined from TTVs due to the smaller transit depth of Kepler-105c, which leads to larger transit midpoint uncertainties for Kepler-105c.

\begin{figure*}%
    \includegraphics[width=0.5\textwidth]{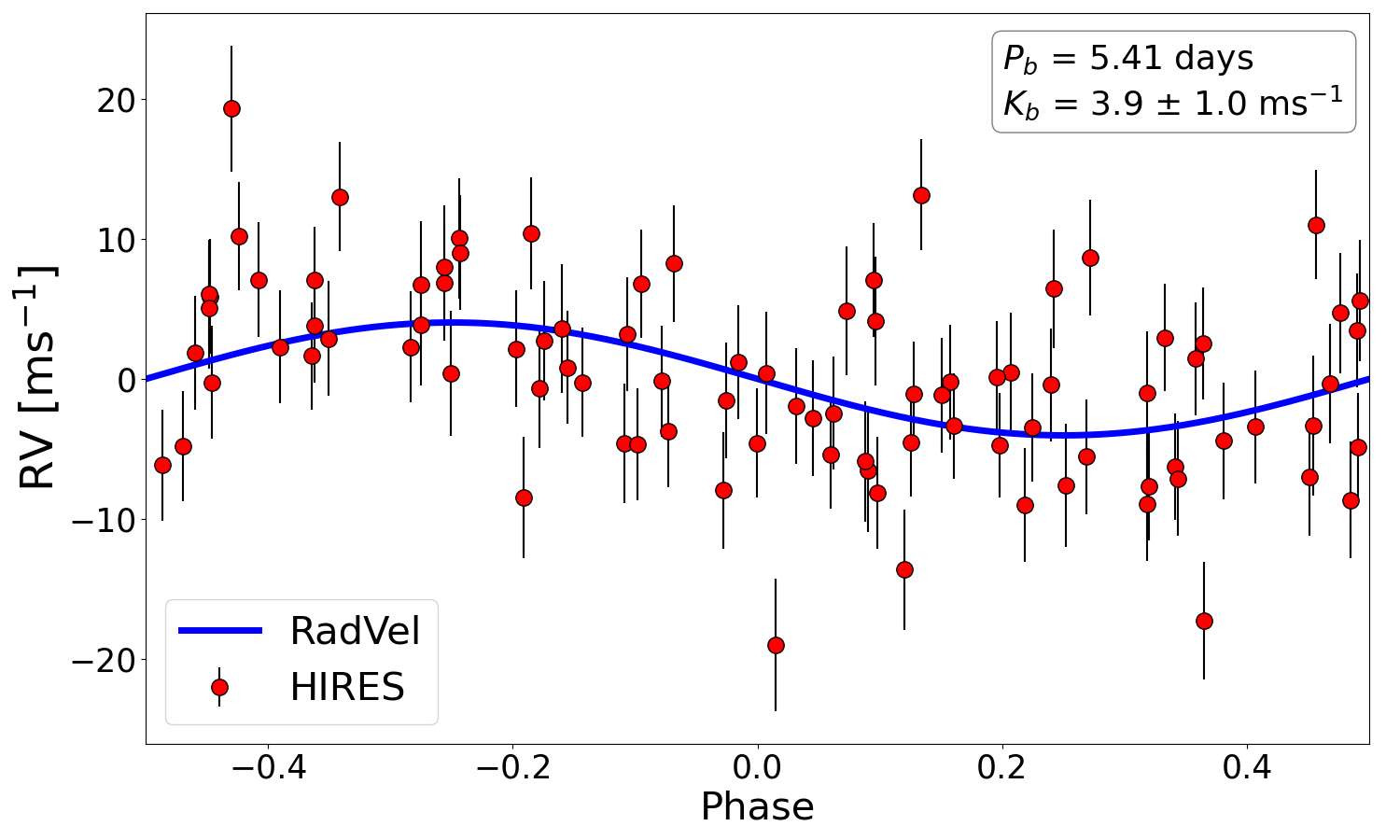}%
     \includegraphics[width=0.5\textwidth]{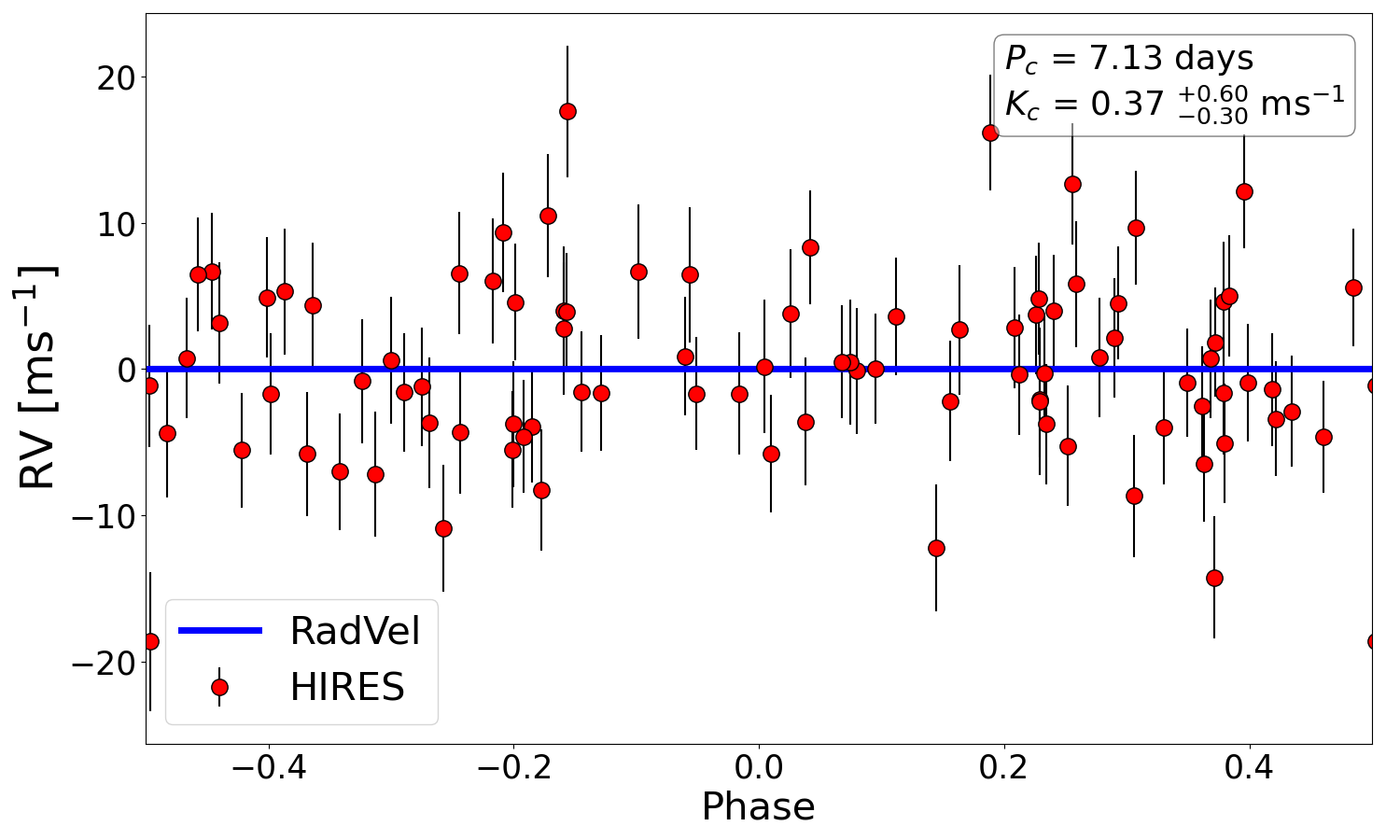}%
    \caption{Phase-folded RVs (red) and the \texttt{RadVel} \citep{2018PASP..130d4504F} fit to the RVs (blue) for Kepler-105b (left) and Kepler-105c (right). The \texttt{RadVel} two-planet fit without a GP yielded a 4$\sigma$ detection of Kepler-105b ($10.7 \pm 2.8$ M$_\oplus$) but did not strongly detect Kepler-105c, despite imposing a planetary signal at its known orbital period. Thus, we can only place a 95\% upper limit of 4.6 M$_\oplus $ on the mass of Kepler-105c based on only the RVs.}%
    \label{fig:example}%
    \end{figure*}
    
\section{Mass From RVs}
\label{Section3}
The RV method is a commonly used exoplanet detection technique that involves measuring the Doppler shift in the emitted light of a star caused by the gravitational influence of orbiting planets. In this paper, we measured 92 RV observations of Kepler-105 (Table \ref{table2}) with the High Resolution Echelle Spectrometer (HIRES) on Keck I \citep{1994SPIE.2198..362V}. The observations of Kepler-105 were performed using the C2 Decker with typical exposure times of 1800 seconds (median S/N at 5500 \r{A} = 89 $\mathrm{pix}^{-1}$). The data was processed through the standard HIRES RV data reduction pipeline \citep{2010ApJ...721.1467H}. 

\subsection{Simple Two-Planet Model}
\label{3.1}
Given the challenges in accurately deducing planetary properties from RV data, which is often complicated by the presence of stellar activity, we faced a decision on how to model the RVs. For instance, we could include a Gaussian process (GP) into our RV model to help model the correlated noise from stellar activity (e.g. \citealt{2014MNRAS.443.2517H,2015MNRAS.452.2269R, 2015ApJ...808..127G}). However, Kepler-105 is a low activity star ($\mathrm {log\phantom{!}} R'_{\mathrm{HK}} = -5.19$), so incorporating a correlated noise model may introduce unnecessary free parameters (e.g. \citealt{Blunt2023}). Thus, we chose to model the RVs twice, both with and without a GP, to determine which model produced a more reliable fit to the data. For our first approach, we implemented a simple two-planet Keplerian model using the Radial Velocity Modeling Toolkit \texttt{RadVel} \citep{2018PASP..130d4504F}. We allowed $P$, $\sqrt{e }$ $\mathrm{cos} \omega_*$, $ \sqrt{e }$ $   \mathrm{sin} \omega_* $, $t_0$ and the RV semi-amplitude ($K$) to vary for both Kepler-105b and Kepler-105c. Similar to our TTV model, we used Gaussian priors on $P$ and $t_0$ and uniform priors on $e$, $\omega$, and $K$ (Table \ref{table1}). Additionally, we include two nuisance parameters, jitter ($\sigma_{jit}$) and gamma ($\gamma$), to account for additional astronomical and instrumental noise and the RV offset, respectively. Lastly, we used a Hill stability prior to prevent orbit crossing (Equation \ref{eq2}). 

\subsection{GP model}
\label{3.2}
We also extended the simple two-planet model in \ref{3.1} by including a GP to model correlated noise in the RVs caused by stellar activity. To fully specify a GP, one must define a covariance function, often referred to as a ``kernel'' \citep{2006gpml.book.....R}. Given the quasi-periodic nature of stellar activity \citep{2015MNRAS.452.2269R,2022MNRAS.515.5251N}, we used a quasi-periodic kernel with \texttt{RadVel}:

\begin{equation}
C_{i j}=A^2 \exp \left[-\frac{\left|t_i-t_j\right|^2}{\lambda_e^2}-\frac{\sin ^2\left(\left|t_i-t_j\right| / P_{rot}\right)}{2 \lambda_p^2}\right]
\end{equation}
Here, $i$ and $j$ are indexes of the covariance matrix $C$ and $A$, $\lambda_e$, $P_{rot}$, and $\lambda_p$ are the hyperparameters of our quasi-periodic kernel, representing the GP amplitude, the exponential decay timescale (a proxy for the lifetime of star spots), the stellar rotation period, and the harmonic complexity, respectively. While GPs provide the flexibility to fit complex data sets, they are notorious for overfitting. To mitigate this issue, we imposed physically motivated priors on the GP hyperparameters:

\begin{equation}
\begin{gathered}
0 \leq A \leq \sigma_{RV} \\
0.5P_{rot} \leq \lambda_e \leq 10P_{rot} \\
0 \leq P_{rot} \leq 50.5 \text{ days} \\
0.5 \leq \lambda_p \leq 5
\end{gathered}
\end{equation}
For $A$, this broad prior prevents the overall GP amplitude from exceeding the standard deviation of the RVs. The prior on $P_{rot}$ is set between 0 and the 3$\sigma$ upper bound reported in \citet{2013ApJ...775L..11M}. For the priors on $\lambda_e$ and $\lambda_p$, we follow the recommendations of \citet{2017PhDT.......229R}. 

\subsection{Three-Planet Models}
In addition to the two-planet models described in Sections \ref{3.1} and \ref{3.2}, we also ran \textit{two} different three-planet RV models.  For these models, we adopt the same approaches as Section \ref{3.1} (non-GP) and Section \ref{3.2} (GP), with the addition of the third candidate planet. Specifically, we included the 0.55 R$_\oplus$ candidate planet at 3.43 days, with a Gaussian prior on $P$ and $t_0$ based on \citet{2018ApJS..235...38T}.

\subsection{Model Comparison}
Based on these set-ups, we ran the MCMC code embedded within \texttt{RadVel} for these models to maximize the following log-likelihood function: 

\begin{equation} 
\begin{aligned}
\label{eq3}
\log(\mathcal{L}) = -0.5 \bigg(\sum_{k} \frac{\left(\mathrm{RV}_{k}-\mathrm{RV}_{m, k} (t_k) \right)^{2}}{\sigma_{\mathrm{RV}, k}^{2}} + \log(\sigma_{\mathrm{RV}, k}^{2})\bigg)
\end{aligned}
\end{equation}
where $\mathrm{RV}_{k}$ is the \(k\)-th observed RV measurement,  and $\mathrm{RV}_{m , k} (t_k) $ is the Keplerian-modeled RV at time $t_k$.  $\sigma_{{RV}, k}$ is the uncertainty for the \(k\)-th RV measurement, which is defined as the observational 
uncertainty ($\sigma_k$) added together in quadrature with a jitter ($\sigma_{jit}$) term: $\sqrt{(\sigma_k)^2 +(\sigma_{jit})^2}$. We ran this MCMC algorithm with 50 walkers until convergence. The initial 10\% of steps were discarded as burn-in. To check for convergence, we once again required the PSRF to be less than 1.01 for each parameter. 

Our two-planet non-GP \texttt{RadVel} fit to the RV data yielded a mass of $10.7 \pm 2.8$ M$_\oplus$ for Kepler-105b. We were unable to strongly detect Kepler-105c in the RVs, so we only placed a 95\% upper limit of 4.6 M$_\oplus$ for this planet. In comparison, our two planet fit with a GP yielded masses of $M_b = 10.2 \pm 2.6$ M$_\oplus$ and $M_c < 3.8$ M$_\oplus$ (95\% upper limit). Thus, both models produced similar masses for Kepler-105b and Kepler-105c.  Given that the results were nearly identical, we determined that the added complexity of a GP is not justified for Kepler-105 since the simpler model is strongly favored ($\Delta AIC = -32$). Furthermore, the two different three planet models did not detect the 0.55 R$_\oplus$ candidate planet, placing 95\% upper limits of 4.6 (non-GP) and 4.8 M$_\oplus$ (GP). These models also failed to detect an RV signal for Kepler-105c, only placing 95\% upper limits of 4.6 (non-GP) and 4.7 M$_\oplus$ (GP). Since these models are strongly disfavored to our two-planet non-GP fit ($\Delta AIC > 38 $ for both models) and no additional planetary signals were detected in  the Lomb-Scargle periodogram of the RVs \citep{Lomb,Scargle}, we conclude that the simple two-planet model is the best fit to the RVs of Kepler-105.

\subsection{Why didn't we detect Kepler-105c in the RVs?}
In an effort to understand why Kepler-105c was not detected in the RVs, we conducted an injection-recovery test where we injected a synthetic planetary signal into the RVs based on the posteriors of Kepler-105c from the TTVs. Even with the injected signal, we still do not strongly detect Kepler-105c in the RV data. This is somewhat surprising: Kepler-105c should generate a signal of $2 $\text{ m $\text{s}^{-1}$}$ $ based on its TTV mass ($5.9 \pm 1.4$ M$_\oplus$). According to Equation 7 of \citet{2016PASP..128k4401H}, a $2 $\text{ m $\text{s}^{-1}$}$ $ signal is expected to be detectable with 6$\sigma$ confidence in a sample of 92 RVs. However, it is important to note that the \citet{2016PASP..128k4401H} relation is primarily derived from the RVs of giant planets. This could limit its relevance to smaller planets like Kepler-105b and Kepler-105c, which may explain why we only detected the more massive Kepler-105b with 4$\sigma$ confidence and did not strongly detect Kepler-105c. Other factors, such as the presence of additional planets or unmitigated stellar activity,  may also contribute to our failure to confidently detect Kepler-105c in the RVs. Furthermore, discrepancies between RV and TTV mass estimates are not unprecedented and have been a subject of ongoing study for many years \citep{2014ApJ...783L...6W, 2016MNRAS.457.4384S, 2017ApJ...839L...8M, 2020A&A...634A..43O}. As a result, this discrepancy between the RV- and TTV-determined mass of Kepler-105c merits further scrutiny, both to understand the specific case of Kepler-105c as well as the broader issue of reconciling measured RV and TTV masses.  

\section{Joint Modeling of RVs and TTVs}
\label{Section4}
\begin{table*}
\centering
\caption{Dynamical parameters of Kepler-105b and Kepler-105c from an RV-only fit (\texttt{RadVel}), a TTV-only fit (\texttt{TTVFaster}), and a simultaneous fit to RVs  and TTVs (\texttt{RadVel} and \texttt{TTVFaster}). Planet parameters were derived based on the stellar parameters reported in \citet{2018AJ....156..264F}. We also report the radii of both planets from \citet{2018AJ....156..264F} to compute their densities, although the planet radii were not directly measured in this paper.  It is also worth noting that in \texttt{TTVFaster}, $180^\circ$ was added to the argument of periastron of the planets ($\omega_b$ and $\omega_c$) to address the inconsistency in the modeling of $\omega$ between \texttt{RadVel} and \texttt{TTVFaster}.}

\label{table1}
\begin{tabular}{|c|c|c|c|c|}
\hline
Parameter& RV-only & TTV-only & Joint RV and TTV & Prior\\
\hline

\hline

\hline

$P_b$ (days) 
& 5.412207 $\pm$ 0.000002 & $5.41220324 \pm 0.0000003$ & $5.4122034 \pm 0.0000004$ & Norm(5.412207130,0.000002488)\\
\hline
$e_b$ & $0.05 \pm 0.04$ & $0.01 \pm 0.01$ & $0.02 \pm 0.02$ & Unif(0,1)  \\
\hline
$\omega_b$ ($^\circ$) & $358.6^{+92.8}_{-159.0}$ & $61.2^{+60.0}_{-142.5}$ & $225.0^{+68.4}_{-137.1}$ & Unif(0,360)\\
\hline
${t_{0,b}}$ (BJD) & $2454955.3185 \pm 0.0006 $ & $2454955.3186 \pm 0.0003$ &  $2454955.3186 \pm 0.0002$ & Norm(2454955.318609,0.000536)\\
\hline
$M_b$ (M$_\oplus$) & $10.7 \pm 2.8$  & $9.3^{+4.9}_{-4.6} $  & $10.8 \pm 2.3$ &  Unif(0,50)\\
\hline
$R_b$ (R$_\oplus$) & $ 2.53 \pm 0.07$  & $ 2.53 \pm 0.07$  &  $ 2.53 \pm 0.07$ & - \\
\hline
$\rho_b$ ($\mathrm{g} \phantom{!} \mathrm{cm^{-3}}$) & $3.65 \pm 1.01 $ &  $3.17^{+1.69}_{-1.59}$ &  $3.68 \pm 0.84$ & - \\
\hline
$P_c$ (days) & $7.12594 \pm 0.00001$ & $ 7.12592 \pm 0.00001$ & $ 7.12592 \pm 0.00001$ & Norm(7.125945910,0.000012500) \\
\hline
$e_c$ & $0.04 \pm 0.04$   & $0.02 \pm 0.02$ & $0.02 \pm 0.02$ & Unif(0,1)  \\
\hline
$\omega_c$ ($^\circ$) & $310.9^{+156.4}_{-112.4}$  & $124.3^{+156.1}_{-67.4}$ & $298.9^{+135.7}_{-59.5}$ & Unif(0,360) \\
\hline
${t_{0,c}}$ (BJD) & $2454957.753 \pm 0.0001 $  & $2454957.754 \pm 0.0003 $ & $2454957.753 \pm 0.0001 $ & Norm(2454957.753432,0.001687) \\
\hline
$M_c$ (M$_\oplus$) & $4.6 \textrm{ (95\% Upper Limit}) $ & $5.9 \pm 1.4 $  &  $5.6 \pm 1.2$ & Unif(0,50)\\
\hline
$R_c$ (R$_\oplus$) & $ 1.44 \pm 0.04$ & $ 1.44 \pm 0.04$ &  $ 1.44 \pm 0.04$ & - \\
\hline
$\rho_c$ ($\mathrm{g} \phantom{!} \mathrm{cm^{-3}}$) & $2.31^{+3.54}_{-1.97}$ & $10.9 \pm 2.75 $ & $10.4 \pm 2.39$  & - \\
\hline

\end{tabular}
\end{table*}

In the previous sections, we analyzed the RVs and TTVs separately. The TTVs placed strong constraints on the mass of Kepler-105c but not Kepler-105b, while the opposite was true for the RVs. To obtain precise mass measurements for both planets, we combined these two methods using a joint RV and TTV model. To do this, we used \texttt{TTVFaster} and \texttt{RadVel} to maximize the log-likelihood function

\begin{equation}
\label{eq4}
\begin{aligned}
\log(\mathcal{L}) = -0.5 \bigg(\sum_{i}                       \frac{\left(\mathrm{TT}_{i}-\mathrm{TT}_{m, i}\right)^{2}}{\sigma_{\mathrm{TT}, i}^{2}}+ \\
\sum_{k} \frac{\left(\mathrm{RV}_{k}-\mathrm{RV}_{m, k} (t_k) \right)^{2}}{\sigma_{\mathrm{RV}, k}^{2}}  + \log(\sigma_{\mathrm{RV}, k}^{2})\bigg)
\end{aligned}
\end{equation}

\normalsize
With this set-up, we used the python package \texttt{emcee} to vary the masses, orbital periods, $\sqrt{e}\cos\omega_*$, $\sqrt{e}\sin\omega_*$, and initial transit times for both planets, as well as the nuisance parameters $\gamma$ and $\sigma$. It is important to note that \texttt{TTVFaster} and \texttt{RadVel} use different conventions where the ascending node and value of $\omega_*$ differ by $180^\circ$ \citep{Householder2022}. Here, we adopt the \texttt{RadVel} convention, which uses a $\hat{Z}$ unit vector that points away from the observer and defines the ascending node as the point where the planet pierces the sky plane moving away from the observer. This is opposite to the coordinate system used in \texttt{TTVFaster}, where $\hat{Z}$ points toward the observer and the planet approaches the observer at the ascending node. To account for this difference, we added $180^\circ$ to the value of $\omega_*$ in the \texttt{TTVFaster} component of our model (note that $\texttt{TTVFaster}$ specifically models $\omega_p$, but it is straightforward to convert between $\omega_p$ and $\omega_*$: $\omega_*$ = $\omega_p +180^\circ$). Similarly to our other models, we implemented uniform priors on $e$, $M$,  $\omega$ and Gaussian priors on $p$ and $t_0$. We also used a Gaussian prior on stellar mass as well as a Hill stability prior. We ran this MCMC for $ 8 \times 10^5 $ steps, discarding the first $10^5 $ steps as burn-in. To ensure that our chains converged, we required the PSRF to be less than 1.01 for each parameter in our model. All of the chains met this PSRF threshold. This MCMC model yielded masses of $10.8 \pm 2.3$ M$_\oplus$ and $5.6 \pm 1.2$ M$_\oplus$ for Kepler-105b and Kepler-105c, respectively (Table \ref{table1}). 

\section{Planet Interiors}
\label{Section5}
Using the radius measurements from \citet{2018AJ....156..264F}, we can now plot Kepler-105b and Kepler-105c on a mass-radius diagram (Figure \ref{Fig.3}). While the masses and radii alone cannot reveal the composition of the planets, this figure does provide some insight into their potential compositions. The mass and radius of Kepler-105c are consistent with a rocky planet without a significant gaseous envelope. Kepler-105b, on the other hand, lies above the 100\% rocky composition line, suggesting that Kepler-105b has a substantial volatile envelope. Assuming that Kepler-105b has an Earth-like core mass fraction of $67.5\%$ $\textrm{MgSiO}_3$ and $32.5\%$ $\textrm{Fe}$ \citep{2007ApJ...669.1279S}, the envelope mass fraction of $\textrm{H}_2$-He of Kepler-105b would be between $0.5\%$-$2\%$ \citep{2014ApJ...792....1L}. 

Another possible composition that has been suggested for planets of similar masses and sizes to Kepler-105b is that of a ``water world:'' a rocky planet with hundreds or thousands of kilometers of water, although the existence of such planets remains a topic of debate \citep{2021JGRE..12606639B,2022ApJ...933...63N, 2023ApJ...947L..19R}. If Kepler-105b is a water-world, it likely would have formed beyond the  $\mathrm{H_2O}$ snow line and migrated inward to its present orbit via Type I migration. This scenario is supported by the fact that Kepler-105b and Kepler-105c are near the 4:3 mean motion resonance, as Type I migration is a common mechanism for the formation of planets in mean motion resonances \citep{2012ARA&A..50..211K}. However, this would require Kepler-105c to form beyond the snow line, which is inconsistent with its high density ($\rho_c = 10.4 \pm 2.39 $ g cm$^{-3}$). It may have been possible for Kepler-105c to form in-situ and for Kepler-105b to migrate from beyond the snow-line, but this would necessitate a fast orbit crossing between the planets. Given these challenges, it seems more plausible that Kepler-105b has a $\textrm{H}_2$-He dominate envelope rather than being a water-world, but it is difficult to make a definitive assertion without better observational evidence. Unfortunately, it will be difficult to determine the precise composition of the atmosphere of Kepler-105b, even with potential follow-up observations. Its atmospheric characterization with the \textit{James Webb Space Telescope (JWST)} is not feasible (see Transit Spectroscopy Metric (TSM),  \citealt{2018PASP..130k4401K}), primarily due to the faint magnitude of Kepler-105 (J $\approx$ 11.8).

\newpage
\section{Scenarios for Atmospheric Mass Loss}
\label{Section6}
\subsection{Photoevaporation}
\label{5.1}

\begin{figure*}%
    \centering
    \includegraphics[width=0.99\textwidth]{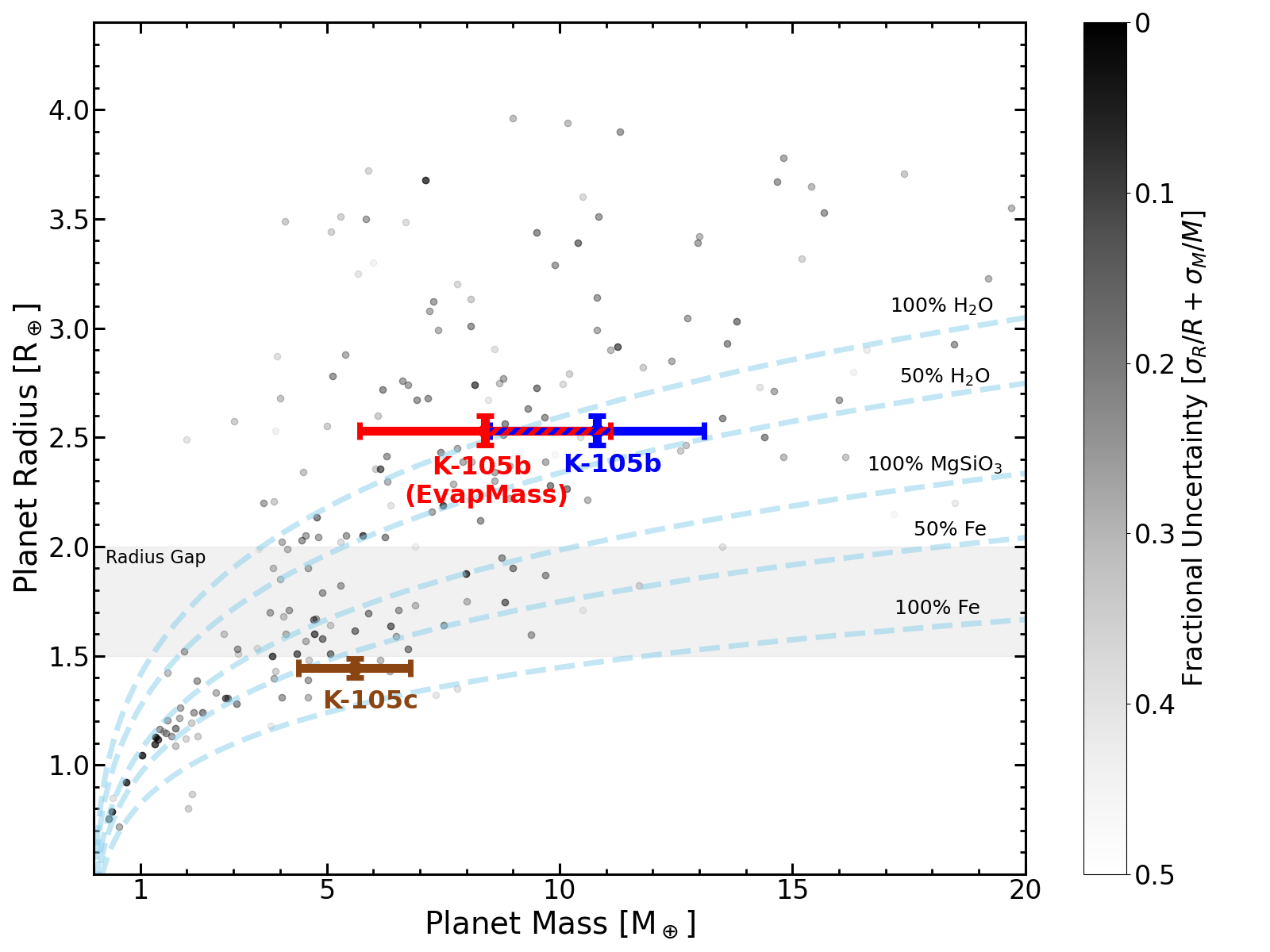}
    
    \caption{The mass-radius relationship for transiting exoplanets with combined fractional mass and radius uncertainties less than 50\% (plotted in grey as a function of fractional uncertainty), based on the NASA Exoplanet Archive (\citealt{2013PASP..125..989A}, queried on September 22, 2022). We also depict the radius gap \citep{2017AJ....154..109F} from $1.5$ R$_\oplus - 2.0 $ R$_\oplus$ (light grey) as well as the planetary compositions (light blue, dashed) from \citet{2019PNAS..116.9723Z}. Additionally, we include the 1$\sigma$ radius \citep{2018AJ....156..264F} and mass measurements of Kepler-105b (blue, $2.53 \pm 0.07R_\oplus$, $10.8 \pm 2.3M_\oplus$) and Kepler-105c (brown, $1.44 \pm 0.04R_\oplus$,  $5.6 \pm 1.2M_\oplus$). We also show the \texttt{\texttt{EvapMass}} \citep{2020ascl.soft11015O} predicted $1\sigma$ minimum mass distribution for Kepler-105b that is consistent with photoevaporation (red, $8.7 \pm 2.4M_\oplus$). If the measured mass of Kepler-105b is greater than the \texttt{\texttt{EvapMass}} prediction, then the difference in gas content between Kepler-105b and Kepler-105c can be explained by photoevaporation. Comparing these distributions, we conclude that the difference in gas envelopes of the Kepler-105 planets is entirely attributable to photoevaporation in 76\% of scenarios (i.e. $M_{b, \mathrm{measured}} > M_{b, \mathrm{predicted}} $)}
    \label{Fig.3}
    
\end{figure*}

Recently, there has been a growing interest in testing photoevaporation models in systems like Kepler-105 where rocky super-Earths are exterior to gaseous sub-Neptunes \citep{2020MNRAS.491.5287O}. Such planetary architectures offer a unique testbed for photoevaporation because the sub-Neptune retained its gaseous envelope despite being subject to more cumulative XUV flux (assuming in situ formation). With the masses from our joint RV and TTV analysis, we can assess if the Kepler-105 planets have a formation history that is consistent with photoevaporation. In this context, ``consistent with photoevaporation'' means that the measured masses and radii of the Kepler-105 planets support the hypothesis that Kepler-105c lost its gaseous envelope due to photoevaporation, while Kepler-105b retained a significant gaseous envelope. To evaluate the validity of this hypothesis, we used the publicly available code \texttt{EvapMass} \citep{2020ascl.soft11015O}. 

We provide a brief outline of the \texttt{EvapMass} numerical procedure, which is more fully described in \citet{2020MNRAS.491.5287O}. \texttt{EvapMass} assumes that both Kepler-105b and Kepler-105c formed in-situ with $\textrm{H}_2$-He envelopes and that Kepler-105c was \textit{just} able to lose its envelope entirely due to photoevaporation, maximizing its atmospheric mass-loss time-scale.  The atmospheric loss time scale, $t_m$, is given by  $t_m = M_{env}/\dot{M}_{env}$ where the equation for the rate of atmospheric mass loss ($\dot{M}_{env}$) is expressed as follows \citep{2020MNRAS.491.5287O}:

\begin{equation}
\dot{M}_{env}=\frac{\eta L_{\mathrm{XUV}} R_p^3}{4 G M_p a_p^2}
\end{equation}
The variables $\eta$, $L_{\mathrm{XUV}}$, and $G$, represent the mass loss efficiency, XUV luminosity of the host star, and the gravitational constant, respectively. Since the Kepler-105 planets have $\sim 99\%$ of their mass in solid materials (section \ref{Section4}), we can assume that $M_p \approx M_{core}$, so the equation for $t_m$ can be written as:

\begin{equation}
\label{eq5}
t_m = \frac{4 G M_p^2 a_p^2 X}{\eta L_{\mathrm{XUV}} R_p^3}
\end{equation}
where $X$ is the envelope mass fraction: $X \equiv M_{{env }} / M_{{core}}$. For Kepler-105c, our goal is to find the envelope mass at which the mass-loss timescale is maximized.  Since we assumed that Kepler-105c formed in situ and that $M_p \approx M_{core}$, $L_{\mathrm{XUV}}$, $M_p$, and $a_p$ are independent of the envelope mass. Thus, we can maximize the following: 

\begin{equation}
t_m \propto \frac{X}{\eta R_p^3}
\end{equation}

\texttt{EvapMass} solves for these dependencies numerically (i.e. computing $X$ as a function of $M$, $a$, and $R$ involves numerically evaluating several integrals) and then compares the mass-loss timescales for the two planets \citep{2020MNRAS.491.5287O}. Since Kepler-105b has a significant gaseous envelope, we require that its atmospheric mass loss timescale is greater than or equal to the maximum atmospheric mass loss timescale of the rocky Kepler-105c. This approach effectively minimizes the mass-loss timescale for Kepler-105b, providing us with a mass lower limit for Kepler-105b that is consistent with photoevaporation. If Kepler-105b had a mass below this value, its mass-loss timescale would be too short to sustain its current gaseous envelope, given that Kepler-105c was stripped of its envelope due to photoevaporation.

\texttt{EvapMass} was specifically designed to compute a minimum mass without measured masses and is often used to report a 95\% limit that the planet mass must be bigger than to be consistent with photoevaporation \citep{2020MNRAS.491.5287O}. However, with the availability of our measured posterior distributions of the Kepler-105 planets, we can adopt a slightly different approach. By randomly selecting samples from these posterior distributions, \texttt{EvapMass} can compute a minimum mass for each sample. We can compare each minimum mass with the corresponding measured mass to determine the percentage of samples where the measured mass of Kepler-105b is greater than its \texttt{EvapMass} predicted minimum mass. A higher percentage of cases where the measured mass is greater the \texttt{EvapMass} predicted minimum mass indicates a higher consistency with photoevaporation.

Since the \texttt{EvapMass} computation depends on both the mass and radius of Kepler-105b and Kepler-105c as well as the properties of their host star (i.e. temperature, mass, radius, age), we evaluated 50,000 randomly drawn samples from these measured distributions. We adopted a value of $1.8$ R$_\oplus$ for the location of the radius gap, a value that is generally accepted for FGK stars, although it can be lower ($\sim 1.5$ R$_\oplus$) for M-dwarfs \citep{2018MNRAS.479.4786V,2021MNRAS.507.2154V}. This selection means that the entire radius distribution of Kepler-105c falls below the radius gap. Our calculations for $\eta$ are based on the hydrodynamical models from \citet{2012MNRAS.425.2931O}. Using \citet{2018AJ....156..264F} for our host star properties and planet radii, combined with our measured mass distribution of Kepler-105c, we computed a minimum mass distribution of $8.7 \pm 2.4$ M$_\oplus$ for Kepler-105b, assuming Kepler-105c was stripped of its envelope due to photoevaporation (Figure \ref{Fig.3}). For each of our 50,000 samples, we compared the measured mass sample of Kepler-105b with the predicted mass sample of Kepler-105b using \texttt{EvapMass}. Our analysis revealed that 76\% of the compared samples were consistent with photoevaporation (i.e. $M_{b, \mathrm{measured}} > M_{b, \mathrm{predicted}} $). Thus, we conclude that it is probable that the difference in gas content of the Kepler-105 planets is consistent with a history of photoevaporation.

For the 24\% of cases that are inconsistent with photoevaporation (i.e. $M_{b, \mathrm{measured}} < M_{b, \mathrm{predicted}} $), we find that these scenarios are also inconsistent with core-powered mass loss in 99\% of cases (details of this procedure can be found in Section \ref{5.2}). In these scenarios, stochastic events such as giant impacts \citep{2015MNRAS.448.1751I, 2019NatAs...3..416B} could explain the differing envelope fractions of the planets. It is also possible that the Kepler-105 planets underwent migration, in which case their present gas envelopes need not be consistent with in-situ mass loss predictions.

\subsection{Core-Powered Mass Loss}
\label{5.2}
Since we tested the viability of photoevaporation to explain the difference in gas content between Kepler-105b and Kepler-105c, it is natural to explore another frequently cited mechanism for the radius gap: core-powered mass loss. Core-powered mass loss relies on the internal heat from a planet's core and the thermal radiation from the host star to drive the evaporation of its atmosphere \citep{2018MNRAS.476..759G}. Rather than conduct a full numerical procedure like we did for photoevaporation, we follow the simpler approach of \citet{2020AJ....160....3C}. Specifically, we required the timescale for core-powered atmospheric mass loss for Kepler-105b to be greater than or equal to that of Kepler-105c. This condition provides the following constraint on planetary parameters (derived in Appendix B of \citealt{2020AJ....160....3C}): 
\begin{equation}
\label{eq7}
\begin{aligned}
1 & \leq\left(\frac{M_{\mathrm{core},b}}{M_{\mathrm{core},c}}\right)\left(\frac{T_{\mathrm{eq},b}}{T_{\mathrm{eq},c}}\right)^{-3 / 2} \\
& \times \exp \left[c^{\prime}\left(\frac{M_{\mathrm{core},b}}{T_{\mathrm{eq},b } R_{b}}-\frac{M_{\mathrm{core},c}}{T_{\mathrm{eq},c } R_{c}}\right)\right]
\end{aligned}
\end{equation}
where $M_{\mathrm{core}} \approx M_p$, $T_\mathrm{eq}$ is the equilibrium temperature of the planet and $c^{\prime}$ is a constant: $\sim 10^4$ K R$_\odot$ ${\mathrm{M}_\odot}^{-1}$. We use host star properties to compute $T_\mathrm{eq}$ for both planets:

\begin{equation}
T_{\mathrm{eq}} = T_{\mathrm{eff}} \sqrt{\frac{R_*}{2a}} (1 - A_B)^{1/4}
\end{equation}
where $T_{\mathrm{eff}}$ and $R_*$ are the temperature and radius of the star, and $A_B$ is the Bond albedo. Assuming Gaussian distributions for $T_\mathrm{eff}$ and $R_*$ ($5933\pm60$ K, $1.03\pm{0.02}$ R$_\odot$) based on \citet{2018AJ....156..264F} and choosing a Bond albedo of 0.3 for both planets, we compute $T_{\mathrm{eq},b} = 1076 \pm 15$ K and $T_{\mathrm{eq},c} = 981 \pm 13$ K. When we apply Equation \ref{eq7} to these equilibrium temperatures and the mass and radius distributions of Kepler-105b and Kepler-105c, we find that these planets satisfy the condition for core-powered mass loss (Equation \ref{eq7}) in 48\% of scenarios. 

\subsection{Varying Bond Albedo}
While our analysis suggests that core-powered mass loss is a plausible explanation for the atmospheric differences in the Kepler-105 planets, it is important to consider the role of Bond albedo in our computation. For instance, if we use a Bond albedo for Kepler-105c that is similar to Venus ($A_b = 0.8$) instead of 0.3, $T_{{\mathrm{eq},c}} = 717 \pm 10$ K. With this single alteration, the consistency of these planets with core-powered mass loss decreases from 48\% to $12\%$. Conversely, if we instead change the Bond albedo of Kepler-105b to 0.8, the consistency increases to 86\%. Thus, while our analysis suggests that core-powered mass loss could potentially explain the differences in gas content between Kepler-105b and Kepler-105c, better measurements of $T_\mathrm{eq}$ or $A_b$ will be necessary for a more definitive assessment. 

We also explored the implications of varying the Bond albedo on the photoevaporation models in Section \ref{5.1}. \texttt{EvapMass} assumes a Bond albedo of 0 for both planets when computing their equilibrium temperature. We found that setting both $A_b$ and $A_c$ to 0.3 for the calculation of $T_{eq}$ resulted in 83\% consistency with photoevaporation. Altering these values to $A_b = 0.3$, $A_c = 0.8$ and $A_b = 0.8$, $A_c = 0.3$ led to slightly different consistencies of 81\% and 93\%, respectively. Since the equilibrium temperature is essentially a proxy for stellar flux in \texttt{EvapMass}, we can also modify the atmosphere's response by varying the opacity, $\kappa$, given by the following: \begin{equation}
\label{eql}
\begin{aligned}
\kappa = \kappa_0 P^{\alpha} T^{\beta}
\end{aligned}
\end{equation}
Here, $\kappa_0$ is the opacity constant and $\alpha$ and $\beta$ describe the pressure (P) and temperature (T) dependence of opacity. By default, \texttt{EvapMass} sets $\kappa_0 = 10^{-7.32}$, $\alpha = 0.68$, and $\beta = 0.45$, where pressure and temperature are expressed in cgs units \citep{2010ApJ...712..974R}. We varied $\kappa_0$ by an order of magnitude (i.e. $ \kappa_0 = 10^{-6.32}, \kappa_0 = 10^{-8.32} $). For these scenarios, the consistency with the photoevaporation model remained 76\%. Thus, the photoevaporation models are less sensitive to changes in Bond albedo compared to core-powered mass loss models. This result aligns with findings from \citet{2012MNRAS.425.2931O}, which demonstrate that photoevaporation mass-loss rates are not highly sensitive to variations in the underlying planetary atmospheric temperature.

Interestingly, systems like Kepler-105 present an opportunity to indirectly constrain the Bond albedo for sub-Neptunes and super-Earths. By jointly modeling photoevaporation and core-powered mass loss in systems like Kepler-105, it may be possible to identify the range of Bond albedos that would allow Kepler-105b to sustain its envelope given that Kepler-105c lost its envelope. This approach could provide us with some of the first Bond albedo constraints for smaller planets, since Bond albedo can typically only be constrained for larger planets with detectable secondary eclipses.

\section{Summary and Discussion}
\label{Section7}
In this paper, we investigated the unusual architecture of the Kepler-105 planetary system, with two planets straddling the exoplanet radius gap in an ideal way for testing photoevaporation. By combining precise radial velocity measurements from HIRES  on Keck I with transit timing variations acquired from the \textit{Kepler Space Telescope} during Q1-Q17, we measured masses of 10.8 $\pm$ 2.3 M$_\oplus$ ($ \rho_b = 3.68 \pm 0.84$  g cm$^{-3}$) and $5.6\pm1.2$ M$_\oplus $ ($\rho_c = 10.4 \pm 2.39 $ g cm$^{-3}$) for Kepler-105b and Kepler-105c, respectively. Our numerical mass predictions with \texttt{EvapMass}  suggest that in 76\% of scenarios, the difference in gas envelope content between Kepler-105b and Kepler-105c can be explained by photoevaporation (i.e. $M_{b, \mathrm{measured}} > M_{b, \mathrm{predicted}} $). However, we acknowledge that alternative mechanisms, such as core-powered mass loss, cannot be definitively ruled out at this stage and warrant further investigation. Furthermore, our mass measurements reveal a $\sim$ 2$\sigma$ mass difference between the cores of Kepler-105b and Kepler-105c. While photoevaporation sculpts the gas envelopes of exoplanets, it does not generate differences in the mass of solid materials, leading to an unresolved question: what mechanism produced the difference in solid mass between Kepler-105b and Kepler-105c? Further investigations into the formation and evolution of Kepler-105b and Kepler-105c will be required to determine the underlying mechanisms responsible for the origin of these planets. 

\section*{Acknowledgements}
We thank the anonymous referee whose insights and suggestions significantly enhanced the quality of this manuscript.

This material is based on work supported by the National Science Foundation REU Program (grant no. 2050527). AH  thanks Beatriz Campos Estrada, Greg Laughlin, Andrew W. Mayo, and the Astroweiss group for useful conversations and feedback. AH also thanks Jason Rowe for generously sharing the transit times used in this paper. We are also grateful to Miki Nakajima for her contributions to the proposal that enabled the acquisition of the RV data presented in this work. 

L.M.W. acknowledges support from the NASA-Keck Key Strategic Mission Support program (grant no. 80NSSC19K1475) and the NASA Exoplanet Research Program (grant no. 80NSSC23K0269). R.A.R. is supported by the NSF Graduate Research Fellowship, grant No. DGE 1745301. J.M.A.M. acknowledges support from the National Science Foundation Graduate Research Fellowship Program under Grant No. DGE-1842400 and from NASA’S Interdisciplinary Consortia for Astrobiology Research (NNH19ZDA001N-ICAR) under award number 19-ICAR19\_2-0041. This work was supported by a NASA Keck PI Data Award, administered by the NASA Exoplanet Science Institute. Data presented herein were obtained at the W. M. Keck Observatory from telescope time allocated to (1) the University of Hawai`i, and (2) the National Aeronautics and Space Administration through the agency's scientific partnership with the California Institute of Technology and the University of California. The Observatory was made possible by the generous financial support of the W. M. Keck Foundation.

The authors also wish to recognize and acknowledge the very significant cultural role and reverence that the summit of Maunakea has always had within the indigenous Hawaiian community. We are most fortunate to have the opportunity to conduct observations from this mountain.
\vspace{5mm}
\facilities{\textit{Kepler}, Keck-HIRES}

\software{
\texttt{RadVel} \citep{2018PASP..130d4504F},
\texttt{TTVFaster} \citep{2016ascl.soft04012A},
\texttt{TTVFast} \citep{2014ApJ...787..132D}, \texttt{NumPy} \citep{numpy}, \texttt{Matplotlib} \citep{matplotlib}, \texttt{Pandas} \citep{pandas}.
}
\clearpage

\bibliography{sample631}{}
\bibliographystyle{aasjournal}

\clearpage
\centering
\begin{longtable}{|c|c|c|c|c|}
\caption{Kepler-105 RV observations and activity indicators (HIRES)}

\label{table2} \\ 
\hline 

BJD & RV ($\text{m $\text{s}^{-1}$}$) &  $\sigma_{RV}$ ($\text{m $\text{s}^{-1}$}$) & $S_{HK}$ & $\sigma_{S_{HK}}$ \\
\hline
\hline
2457197.948219 & -17.85 &    3.03 &  0.146 &      0.001 \\ \hline
2457200.991239 &  -4.32 &    2.74 & 0.1402 &      0.001 \\ \hline
2457202.062219 &  -5.11 &    2.57 & 0.1422 &      0.001 \\ \hline
2457204.037027 &   2.88 &    2.83 & 0.1406 &      0.001 \\ \hline
 2457222.06666 &  -1.24 &    3.12 & 0.1384 &      0.001 \\ \hline
2457229.095178 & -14.22 &    3.20 & 0.1347 &      0.001 \\ \hline
2457229.900375 &  -6.14 &    2.94 & 0.1377 &      0.001 \\ \hline
2457236.943638 &  18.74 &    3.43 & 0.1225 &      0.001 \\ \hline
2457245.979016 &  -1.00 &    2.79 & 0.1357 &      0.001 \\ \hline
 2457254.93196 &   2.63 &    2.85 & 0.1338 &      0.001 \\ \hline
2457255.995104 &  -7.12 &    3.34 & 0.1256 &      0.001 \\ \hline
2457262.890422 &   1.94 &    2.79 & 0.1382 &      0.001 \\ \hline
2457265.012706 &   9.46 &    3.16 & 0.1264 &      0.001 \\ \hline
2458627.914381 &   9.63 &    2.61 & 0.1498 &      0.001 \\ \hline
2458679.988635 &  -5.32 &    2.33 & 0.1449 &      0.001 \\ \hline
2458714.842907 &   6.50 &    2.43 & 0.1295 &      0.001 \\ \hline
2458722.907216 &  -1.66 &    2.35 & 0.1288 &      0.001 \\ \hline
2458765.847403 &  -3.04 &    2.81 & 0.1314 &      0.001 \\ \hline
2458776.858662 &   3.57 &    3.60 & 0.1236 &      0.001 \\ \hline
2458795.779808 &   6.51 &    2.93 & 0.1271 &      0.001 \\ \hline
2458999.960787 &  -9.50 &    2.91 & 0.1357 &      0.001 \\ \hline
2459003.060371 &  -5.21 &    3.14 & 0.1347 &      0.001 \\ \hline
2459003.971311 &  -5.99 &    2.54 & 0.1344 &      0.001 \\ \hline
2459006.950713 &   1.71 &    2.84 &  0.138 &      0.001 \\ \hline
2459007.993173 &   1.57 &    3.02 & 0.1358 &      0.001 \\ \hline
2459010.998866 &   0.85 &    2.77 & 0.1375 &      0.001 \\ \hline
2459011.985675 &   1.31 &    2.87 & 0.1374 &      0.001 \\ \hline
2459012.983941 &   3.27 &    3.18 & 0.1363 &      0.001 \\ \hline
2459014.048065 &  -0.73 &    2.72 &  0.134 &      0.001 \\ \hline
2459016.916247 &  -7.59 &    3.09 & 0.1322 &      0.001 \\ \hline
2459024.926903 &   7.67 &    2.99 & 0.1313 &      0.001 \\ \hline
2459028.860562 &  12.46 &    2.64 & 0.1362 &      0.001 \\ \hline
2459030.954948 &  -3.39 &    2.98 & 0.1313 &      0.001 \\ \hline
2459035.972648 &  -8.55 &    3.02 &  0.133 &      0.001 \\ \hline
2459040.000646 &   1.72 &    2.72 & 0.1323 &      0.001 \\ \hline
2459071.053519 &  -3.92 &    4.11 & 0.1172 &      0.001 \\ \hline
2459078.03372 &   6.28 &    2.99 & 0.1297 &      0.001 \\ \hline
2459088.889249 &  -0.20 &    3.41 &  0.114 &      0.001 \\ \hline
2459091.966282 &  -1.57 &    3.34 & 0.1187 &      0.001 \\ \hline
2459101.920248 &  -0.80 &    2.94 & 0.1241 &      0.001 \\ \hline
2459114.882941 &   5.32 &    2.90 & 0.1251 &      0.001 \\ \hline
2459117.819583 &   6.51 &    2.86 & 0.1273 &      0.001 \\ \hline
2459119.838001 &  -0.91 &    3.16 & 0.1221 &      0.001 \\ \hline
2459120.820908 &   2.31 &    2.93 & 0.1258 &      0.001 \\ \hline
2459121.879195 &   0.25 &    2.86 & 0.1224 &      0.001 \\ \hline
2459122.889507 &  -2.53 &    2.99 & 0.1248 &      0.001 \\ \hline
2459123.838387 &  -0.13 &    3.20 & 0.1245 &      0.001 \\ \hline
2459153.804096 &   7.43 &    3.34 & 0.1165 &      0.001 \\ \hline
2459362.071766 &  -4.08 &    2.57 & 0.1505 &      0.001 \\ \hline
2459373.881189 &  -4.01 &    2.83 & 0.1465 &      0.001 \\ \hline
2459376.952544 &  -2.11 &    2.94 & 0.1455 &      0.001 \\ \hline
2459377.907244 &  -1.74 &    2.90 & 0.1439 &      0.001 \\ \hline
2459378.938855 &  -6.85 &    2.47 & 0.1484 &      0.001 \\ \hline
2459379.962735 &  -5.36 &    2.67 & 0.1472 &      0.001 \\ \hline
2459383.035088 &  -8.71 &    2.78 & 0.1404 &      0.001 \\ \hline
2459385.946136 &   1.06 &    2.53 & 0.1446 &      0.001 \\ \hline
2459386.913853 &   9.81 &    2.77 & 0.1459 &      0.001 \\ \hline
2459387.954846 &  -0.16 &    3.25 & 0.1482 &      0.001 \\ \hline
2459388.972634 &  -0.43 &    2.75 & 0.1479 &      0.001 \\ \hline
2459389.978382 &  -5.01 &    3.00 & 0.1461 &      0.001 \\ \hline
2459395.952932 &  -9.22 &    2.99 & 0.1395 &      0.001 \\ \hline
 2459399.92468 &  -9.62 &    2.87 & 0.1422 &      0.001 \\ \hline
 2459405.02193 &  -3.94 &    2.42 & 0.1408 &      0.001 \\ \hline
2459406.013179 &  -7.70 &    2.91 & 0.1415 &      0.001 \\ \hline
 2459406.93291 &  -6.74 &    2.71 & 0.1419 &      0.001 \\ \hline
2459409.033993 &  -5.26 &    2.76 & 0.1408 &      0.001 \\ \hline
2459410.042219 &  -6.46 &    3.20 & 0.1403 &      0.001 \\ \hline
2459415.055976 & -19.57 &    3.76 & 0.1379 &      0.001 \\ \hline
2459422.966883 &   4.13 &    3.20 & 0.1385 &      0.001 \\ \hline
2459435.851129 &  -0.82 &    2.62 & 0.1348 &      0.001 \\ \hline
2459441.002974 &  -9.05 &    3.24 & 0.1315 &      0.001 \\ \hline
2459441.953438 &   0.62 &    2.90 & 0.1377 &      0.001 \\ \hline
2459443.838183 &   2.39 &    2.57 & 0.1351 &      0.001 \\ \hline
2459445.03636 &  -0.84 &    2.81 & 0.1365 &      0.001 \\ \hline
2459445.96299 &   6.17 &    3.46 & 0.1362 &      0.001 \\ \hline
2459448.921949 &   8.09 &    2.96 & 0.1356 &      0.001 \\ \hline
2459449.919675 &  10.45 &    2.60 & 0.1343 &      0.001 \\ \hline
2459450.90239 &   3.24 &    2.91 & 0.1392 &      0.001 \\ \hline
2459451.919987 &   2.17 &    3.14 & 0.1354 &      0.001 \\ \hline
2459452.858454 &  -5.16 &    2.59 & 0.1338 &      0.001 \\ \hline
2459455.849691 &   5.52 &    2.53 & 0.1369 &      0.001 \\ \hline
2459456.959194 &   8.45 &    2.93 & 0.1362 &      0.001 \\ \hline
2459469.823537 &  12.58 &    2.70 & 0.1363 &      0.001 \\ \hline
2459470.831935 &  -8.23 &    2.56 & 0.1369 &      0.001 \\ \hline
2459471.755624 &  -5.45 &    2.53 &  0.137 &      0.001 \\ \hline
2459474.903543 &   4.31 &    3.59 & 0.1331 &      0.001 \\ \hline
2459475.876222 &  -8.19 &    3.30 & 0.1285 &      0.001 \\ \hline
2459482.91076 &   4.51 &    3.30 &  0.132 &      0.001 \\ \hline
2459484.819239 &   6.21 &    2.56 & 0.1373 &      0.001 \\ \hline
2459489.884357 &   3.00 &    3.56 &  0.134 &      0.001 \\ \hline
2459498.826307 &   5.04 &    3.21 & 0.1404 &      0.001 \\ \hline
2459502.881921 &   5.88 &    3.09 & 0.1353 &      0.001 \\ \hline
2459503.837569 & -11.50 &    2.93 & 0.1359 &      0.001 \\ \hline

\end{longtable}

\end{document}